\documentclass{article}

\usepackage{arxiv}
\usepackage[utf8]{inputenc} 
\usepackage[T1]{fontenc}    
\usepackage{hyperref}       
\usepackage{url}            
\usepackage{booktabs}       
\usepackage{amsfonts}       
\usepackage{nicefrac}       
\usepackage{microtype}      
\usepackage{graphicx}
\usepackage{natbib}
\usepackage{doi}
\usepackage{program}
\usepackage{listings}
\usepackage{color}

\definecolor{dkgreen}{rgb}{0,0.6,0}
\definecolor{gray}{rgb}{0.5,0.5,0.5}
\definecolor{mauve}{rgb}{0.58,0,0.82}

\title{A Review on Analysis and Visualization Methods for Biclustering}

\author{ \href{https://orcid.org/0000-0002-9698-4360}{\includegraphics[scale=0.06]{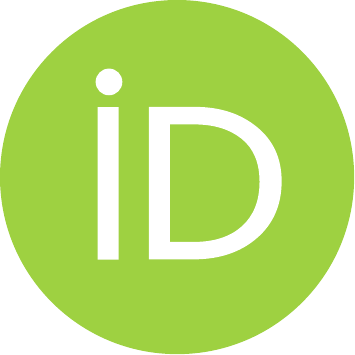}\hspace{1mm}Melih S\"{o}zdinler}$^{1,2}$ \\ 
	$^{1}$ Huawei Turkey R\&D Center, \\
	$^{2}$ Department of Computer Engineering \\
	Bogazi\c{c}i University ,Bebek,Istanbul 34342 Turkey \\
	\texttt{melih.sozdinler@boun.edu.tr}\thanks{Melih S\"{o}zdinler is supported by The Scientific and Technological Research Council of Turkey(TUBITAK)[BIDEB-2211].} 
}

\date{}

\renewcommand{\headeright}{Technical Survey}

\renewcommand{\shorttitle}{A Review on Analysis and Visualization Methods for Biclustering}

\hypersetup{
pdftitle={A Review on Analysis and Visualization Methods for Biclustering},
pdfsubject={q-bio.NC, q-bio.QM},
pdfauthor={Melih Sozdinler},
pdfkeywords={Bicluster, Visualization, Survey},
}

\begin{document}

\maketitle
\begin{abstract}
Recently, biclustering is one of the hot topics in bioinformatics and takes
the attention of authors from several different disciplines. Hence, many different
methodologies from a variety of disciplines are proposed as a solution to
the biclustering problem. As a consequence of this issue, a variety
of solutions makes it harder to evaluate the proposed methods.
With this review paper, we are aimed to discuss both analysis and
visualization of biclustering as a guide for the comparisons
between brand new and existing biclustering algorithms.
Additionally, we concentrate on the tools that provide
visualizations with accompanied analysis techniques.
Through the paper, we give several references
that is also a short review of the state of the art for the ones
who will pursue research on biclustering.
The Paper outline is as follows; we first give the visualization and analysis
methods, then we evaluate each proposed tools
with the visualization contribution and analysis options,
finally, we discuss future directions for
biclustering and we propose standards for future work.
\end{abstract}


\section{Introduction}

The idea of biclustering is first introduced by
Hartigan~\cite{hartigan72}, is also the specialized
version of {\it clustering} problem.
In essence, clustering refers to the process of organizing a set of input vectors into clusters based on specified similarity
with respect to some predefined distance measure~\cite{CM09}.
In some cases having the relaxed types of clusters
that can group input vectors both horizontally and vertically or
in other words, clustering both features and samples are more appreciated.
This special instance of clustering, named as {\it Biclustering}.
In general, clustering can have a resulting intuition that can be
valuable in terms of a global perspective. The local perspectives
and correlations are somehow disregarded by clustering algorithms
unless the Principal Component Analysis or other dimensionality reduction
methodologies are applied.
Using biclustering, we can give both global and local perspectives
by arranging the size of biclusters.

Biclustering is still a hot topic
because of many opportunities to
extend the problem and takes the attention from
sub-disciplines of computer science, mathematics
and statistics. This leads to several application areas
such as data mining, pattern recognition,
micro-array analysis, drug activity analysis, and      
motif detection ~\cite{BCKY02,TSS02,MK03,KBCG03,BIB03,AH06,PBZW06,MA09,CC00,CM09,KP07,
citeulike:6076707,citeulike:1356776,citeulike:2762934,bivusu2,
19208127,nsNMF,citeulike:2714170,citeulike:5649168}.
Particularly, biclustering problem turns into
the optimization problem in several different types of research,
concentrate on solving the specified optimization
problem. In most cases, the problem is
NP-Hard and heuristic solutions are needed. Many
heuristic opportunities make the topic up-to-date.
Refer to these survey papers~\cite{MO04,Tanay2004,citeulike:2402119}
for more details.

Although the topic is hot, both the visualization
and the analysis of the results of biclustering
needs more attention. In this review paper,
we will give the analysis and visualization methods
of existing tools. Then, we compare
pros and cons of these tools briefly. Since
biclustering is important for both finding local and
global perspective of applied biological data, we
should be able to verify the success
of biclustering on applied data using both
computational analyses, visualizations, and
biological validations. In the literature,
we see the congestion that there is already
algorithm explosion in 10 years starting with
Cheng and Church~\cite{CC00}, but still
there is a lack of unique comparison methodologies
rather than scoring comparisons like {\it P-value}
and {\it H-value}. In some cases,
the proposed method can be application-specific
such as applying on different
paradigms of bioinformatics. We distinguish these methods. In the case
when it is just another algorithm for biclustering on
gene expression data,
we believe that a fair sophisticated comparison
methodologies for validation using both
existing biological data and metrics are vital.
We also suggest that existing specialized visualization
methods could also be beneficial as a supporting
evidence of the quality of biclusters. For this purpose,
we will discuss the main analysis and visualization methods
throughout the paper. We expect that this review should
be useful for the forthcoming studies.

Recently, several tools for biclustering that
have embedded supports for the visualization models
and analysis methods. The main visualization
models are Heatmap Representation~\cite{bicat,biggest,bivoc,bicoverlapper}
and Parallel Coordinate Plots~\cite{bicat,biggest,bivoc,bicoverlapper,bivusu}.
Rather than these, there are also specialized visualization models. These are
Force Directed Layout Model~\cite{bicoverlapper}, Bubble Map Model~\cite{bicoverlapper},
Mountain Map Model~\cite{gcloto},
Enrichment Tree Visualization~\cite{biggest}, Integrated Visualization of
Biclusters Verifying with Protein-Protein Interactions~\cite{AES10},
Advanced Heatmap Representation~\cite{bivoc} and
Modified Parallel Coordinate Visualizations~\cite{biggest,bivusu,bicoverlapper}.
BiCluster Viewers~\cite{BiClusterViewer} is also be applied to highlight detected biclusters generated from the original data set by using heatmaps and parallel coordinate plots as visualization methods. BiCFlows~\cite{bicflows}
also provided a novel approach to the visualization of bipartite graphs where well fits into bicluster visualization too. The tool allows for multi-scale exploration through the hierarchical aggregation of nodes and edges using biclustering in the linked lists. Different than all these visualization methods, Forestogram~\cite{Forestogram} specifically considers Hierarchical Biclusters and they provide a 3D method inspired by dendrograms to visualize biclusters.

At ~\cite{Streit2014FurbyFF}, the Furby is another interactive visualization technique for analyzing biclustering results. They have twofold contribution claims. The first one is a high-level view of the overall results. And, that shows bicluster shared rows and columns with visualization. The second contribution is to provide heatmaps and bar charts and enable analysts to interactively set the thresholds that transform the fuzzy (soft) clustering into hard clusters.

Furthermore, several analysis methods are also used in literature. These are
Significantly Enriched Gene Ontology Categories Analysis(FuncAssociate) or Enrichment
Analysis or P-Value~\cite{funcassociate,citeulike:136570,KP06,CM09},
H-value Scoring or Mean Squared Residue Score(MSRS)\cite{CC00},
Hv-value Scoring~\cite{KP07},
Average Correlation Value(ACV)~\cite{LW06},
Pearson Correlation Coefficient~\cite{citeulike:5739089},
validation with previously known results~\cite{XL2007,MK03},
and validation using biological networks~\cite{PBZW06,CM09}.

In this review paper, we first discuss the methodology
of each visualization and analysis method.
Then, we overview proposed tools and their pros and cons.
Finally, we talk about future directions and discussions.
      
\section{Visualization and Analysis Methods}

In the introduction, we review the recent literature
of the visualization and analysis methods.
In terms of visualization methodologies, there are several different approaches.
We divide these into traditional ones and specialized ones.
Traditional ones have their own root as an application of existing
models to the visualization of biclusters. On the other hand, specialized ones
have different properties that are the extensions of existing models,
or completely new models. Furthermore, for some visualization approaches,
we observe that there are efforts to give the resulting picture as a whole
system in one layout and supporting sub-layouts with some interactivity.
On the other hand, the remaining approaches consider one by one approach
which means that for each bicluster, they give the corresponding
visualizations and there are no such high-level visuals. Similarly,
we can make a different classification for analysis methods;
analysis with computation and analysis with biological validation.
In several recent studies, analysis of biclusters conveys the
results of the validation using a scoring function. Also, there
are biological validation methods using supplementary biological networks,
existing natural groupings, and experimental data.

In this section, we first mention visualization methods, and then we
review analysis methods.

\subsection{Visualization Methods}

We now review the visualization methods in two
different titles. First, we will show traditional
approaches, next we will demonstrate
specialized ones.

\subsubsection{Traditional Approaches}
      
Heatmap Representation and Parallel Coordinate Visualization
are assumed as traditional hence they introduced
before the biclustering essentially for data visualization.
In general, they are useful because you can obtain basic
knowledge about the biclusters
by looking at the resulting pictures of these approaches.

Heatmaps show us the selected parts of the dimensions and values corresponding
to these selected dimensions of the dataset, also referred to as biclusters. The main
contribution of heatmaps is their contribution to infer information about the bicluster structure.
By looking at the resulting heatmap, we see the structure of a bicluster and
we can predict the type of a bicluster such as {\it all constant},
{\it constant row},{\it constant column}, {\it coherent}... Particularly,
the structure of the bicluster in heatmap depends on the optimization
criteria and scoring metric of the algorithm. Therefore, heatmaps are one of
the basic validation of the results. A Colouring scheme is vital for heatmaps.
Mainly, variations of colors (green-red) are used and each color represents the
interval of values. Additionally, choosing an appropriate interval is
necessary as well in order to evenly distribute the colors.
Heatmaps are included in several tools~\cite{biggest,bicat,
Sharan03clickand,nsNMF,bicoverlapper,bivoc,AES10} and each tools
have some different modifications.
 
Parallel Coordinates(PC) plots are another useful visualization of biclusters.
PC plots mainly introduce visuals for one dimension of data
with respect to other dimensions.
From the perspective of gene expression data, the PC plots represent the genes
under the subset of conditions. Using PC plots, one
can easily detect the bicluster structure and it is easier to
see {\it coherent} biclusters than heatmaps. In PC plots,
if the overall picture has
simultaneous increasing and decreasing plots,
this means that the corresponding bicluster
has a correlation in itself. Furthermore, PC plots need to have
an appropriate coloring scheme and scaling of plots.
Some approaches also add fuzzy effects on each plot to simplify
the complicated drawings. Also, scaling is vital to show peak points,
intervals of values. Parallel Coordinates are supported in several tools,
makes it fundamental for biclustering tools~\cite{biggest,
Sharan03clickand,bicat,bivusu,bivusu2,bicoverlapper,AES10}.
Note that, some of these tools provide PC plots as
{\it Gene Expression Profiles}~\cite{bicat,biggest}
which is the non-scaled version of PC plots. We will
not distinguish these similar approaches through the review.

Indeed, these two approaches are simple but not enough to
see the whole picture and overall quality in both
computationally and biologically. Overall,
they are useful but more specialized approaches
are required.
 
\subsubsection{Specialized Approaches}

The general property of specialized approaches is to show
the biological relevance of biclusters and to give
the general picture of the results while proposing
a new idea or extending the existing methods.
We review these methodologies in this subsection.

The first visualization approach is based on Force Directed
Layout~\cite{bicoverlapper}. The authors' claim is
to unravel trends and to highlight relevant genes and conditions
using both visual approaches and complementing biological and
statistical analysis. Hence, end-users
may have a chance to explore the results quickly and interactively.
The visualization technique obtains its root
from force-directed layouts that is represented as flexible
overlapped groups of genes and conditions.
The model is integrated with Heatmaps and
Parallel Coordinate Plots and its advantage
is the availability of the extension of its visualization methodology with
biological relevance using transcriptional modules.
Moreover, their proposed model can show several biclusters
at once that also combines the overall view
with the traditional approaches.
Additionally, in their model,
if the nodes are connected, a spring force keeps the nodes closer.
Also, there is an expansion force that pushes
every pair of nodes whether connected or not~\cite{force}.
Their claim is that nodes in the same biclusters are closer and
the remaining nodes belonging to different biclusters are separated.
To represent the bicluster as a graph, they form a complete
graph for each bicluster. For overall visibility, they do not show
all the edges and nodes. Instead, they show the hull of each
bicluster with some transparency. The transparency of this hull
is useful to show overlapping biclusters and to increase visibility.

The second approach is a tree-based visualization method that gives
biological relevance to biclusters. The method makes a connection
between functional categories of an organism in Gene Ontology~\cite{citeulike:5044135}
and the corresponding biclusters. It is proposed by \cite{biggest}.
In their method, they form a Gene Ontology(GO) category
tree for each bicluster. The tree is in hierarchical
layout and specification increases at each lower level of
the layout. Next, they calculate Bonferroni corrected
P-value overall category nodes of the constructed tree.
They color the GO categories on the behalf of lower p-values.
In addition to that, they have also different colors to show GO main
categories;{\it cellular component},
{\it molecular function} and {\it biological process}.
The intensity of each color changes according to
the calculated P-value.
The main contribution of this approach is to
integrate the biclustering concept with the fair
scoring of P-values. Rather than this approach,
FuncAssociate~\cite{funcassociate} and several other tools
have to support to the calculation of P-values. With \cite{biggest}
approach, we observe these results
with a well-readable picture rather than text and
we are able to see the hierarchy of GO categories inside the visual.
The origin of the approach depends on the hierarchical layout
using a well-known graph drawing tool
named graphviz~\cite{graphviz}. Since graphviz provides
nice visualizations, the picture of the layout is
well readable and useful to detect enriched categories.
Additionally, a similar approach is proposed
as in the histogram format~\cite{Sharan03clickand}.
Each histogram shows both the significant transcription
factors and GO functional enrichment
analysis of bicluster genes. Users can
easily detect the most common transcription factor
or most common category according to the calculated P-value
using the histogram.

Recently, An integrated model for visualizing biclusters from
gene expression data and PPI networks(IntegratedViz)
the tool is introduced~\cite{AES10}.
They proposed an approach that
integrates Protein-Protein Interaction(PPI)
networks and biclusters.
Their method has one central graph that represents
each bicluster as a node and each node in
the main graph has a peripheral graph
that corresponds to the sub-network
formed by genes of the bicluster.
The biological relevance is maintained by
using these sub-networks of each bicluster and
the edges between nodes of central
graph. The Layout is based on {\it Weighted
Hierarchical Layout}. Authors propose
this layout as an extension to
{\it Unweighted Hierarchical Layouts}
as a new graph drawing approach.
Introducing weights into graphs
give more options to enrich the layout
to demonstrate extra weight information of edges.
Furthermore, comprising the peripheral graphs are done
such that genes of the corresponding
biclusters are extracted from the PPI network and
eventually, using weighted hierarchical layout algorithm
the final layout is obtained. Peripheral
graph edges between genes show the reliability of the interactions.
At the central graph, scores of each bicluster
are calculated and nodes or biclusters in the
the main graph have a size proportional to these scoring
functions. They use three scoring functions;
{\it H-value}~\cite{CC00}, {\it Hv-value}~\cite{KP07}
and Enrichment Ratio similar to {\it P-value}.
Finally, edges in the central graph also show
either the common genes or interacting edges
between two biclusters. The advantage of IntegratedViz
is its integration of PPI networks and biclusters,
to show the biological relevance of each
set of genes of biclusters in global and
local views. With this idea, correlated
biclusters tend to have more interactions
in their corresponding PPI networks.

Another approach in \cite{bivoc},
extends the traditional heatmap visualization.
The proposed idea shows all biclusters on a special type
of heatmap. The methodology extends
the heatmap layout by including multiple
labels of genes and conditions in the resulting heatmap.
So, they are able to show each bicluster while
maintaining the minimum number of repetitions of
labels. They propose a novel algorithm,
based on {\it PQ-Tree}. The algorithm is based on finding an ordering
such that the binary formation of $1$'s at each bicluster
is consecutive. This is called {\it Consecutive One's Property (COP)}.
In the beginning, discretization of bicluster data is
needed in the form of $0$'s and $1$'s. Then, they set-up
{\it PQ-Trees} from each discretized bicluster $M$
and next, its rows are stored in list $L$.
Using {\it REDUCE} operation, they
perform hierarchical clustering to
maintain COP property. Next, using
the {\it MERGE} operation,
they form resulting {\it PQ-Trees}
as a new list, $L'$ by looking at the similarity score
of column lists $C_T$ and $C_{T^{*}}$ where
$T$ and $T^{*}$ are separate {\it PQ-Trees}.
The similarity score is as follows.

\begin{equation}      
\label{f3}      
\sigma(T,T^{*})= \frac{C_T \cap C_{T^*}}{C_T \cup C_{T^*}}
\end{equation}

$\sigma(T,T^{*})$ is a function to decide merge operation and each
$\sigma$ function results for all pairs are calculated at the beginning
and sorted. Then, they perform {\it REDUCE} operation between
the pairs with the highest $\sigma$. If it fails, {\it MERGE} operation
does not occur. In {\it REDUCE} operation, basically, they are checking
that the restrictions defined by {\it PQ-Tree}, $T$, holds
for $T^{*}$. If it holds, {\it MERGE} operation occurs.
$T$ and $T^{*}$ are deleted from $L$. Merged tree $T_{m}$ is
added to list by upgrading $\sigma$ values.
Finally, when all $\sigma$ values are processed, they give the
final layout as it appears on the set of columns of {\it PQ-Trees}.
This method provides a combinatorial algorithm that holds for
the minimum number of repetitions of labels where they are part
of the original data in the resulting heatmap. Their method works better in
overlapping biclusters and makes it available to show
several biclusters in one heatmap. In several overlapping bicluster cases,
their problem, as they mentioned, is the number of biclusters.
To avoid this problem, they develop a web-based interface that
allows execution and navigation through the web. There is also
another specialized methodology on heatmap visualization in
\cite{20096121}. Although they mainly discuss the
clustering point of view, their proposed toolbox should
be applicable to biclustering. In that approach, they extend the
view of heatmaps into the third dimension using dendrograms and
it is more desirable in such a case to see all biclusters
in one visualization.

In addition to all these mentioned approaches, in \cite{bivusu},
they propose an extended parallel coordinate visualization.
Their approach is to give parallel coordinate plots of a bicluster by
simultaneously drawing with the real data plots and
the bicluster plots of the same conditions.
They are colored with different colors
and bicluster gene plots are more visible to emphasize.
According to this claim, the global view of the
parallel plots is not hidden. This provides a better
understanding of gene plots over a subset of conditions.
Furthermore, in \cite{biggest}, it is not a brand new method,
but they have one screen that includes all plots. They provide
all bicluster plots together to speed up the plots extraction process
and this gives us to look at several results to investigate both local
and global patterns of biclusters. Finally, in ~\cite{bicoverlapper}
they perform some improvements to obtain good scaling and
colorings.

Bubble Map is also another simple approach used in the literature,
but in biclustering, it has a special meaning to
show the projections of Mountain Maps~\cite{gcloto}.
This way, we can represent the bicluster as
2D bubbles with different sizes. Due to the method
suggestion, we accept it as innovative ones.
In the method, each bubble
can have meaning such as correlated patterns inside
the biclusters and higher values in subspaces.
$3D$ version of bubble map is mountain map.
Mountain maps are proposed for cluster
visualization~\cite{gcloto}. In ~\cite{bicoverlapper},
they propose bubble map representation as
a projection of 3D mountains in 2D as bubble maps.
 
In the conclusion of this section, visualization methods
provide us with both local and global perspectives
of biclusters. In recent trends, integrating biclusters with
biological analysis and data becomes more popular and next-generationvisualization tools should include this integration.
We now discuss the existing Analysis Methods.

\subsection{Analysis Methods}

There are several scoring schemes for analysis and we assume that
scoring is a good metric for the computational quality of biclusters.
On the other hand, it is also important to analyze biclusters
with biological knowledge and natural groupings. Therefore, we divide
this subsection into two parts; Analysis with Computation and
Analysis with Biological Validation.

\subsubsection{Analysis with Computation}

We have several scoring functions in order to analyze the
resulting biclusters. For this purpose, we are mainly interested
in the extracted submatrix $A$ of original data using
biclustering. The first scoring method, H-value~\cite{CC00} is found
by calculating residues for each entry in submatrix $A$. Assuming
each bicluster as a submatrix that consist of $I$ rows and $J$ columns,
the residue $R$ of an entry $(i,j)$ is
 
\begin{equation}      
\label{f1}      
R_{I,J}(i,j) = A_{i,j} - A_{Ij} - A_{iJ} + A_{IJ}      
\end{equation}      

where $A_{iJ}$ is the mean of row $i$, $A_{Ij}$ is the mean of column $j$      
and $A_{IJ}$ is the mean of the all $(i,j)$ pairs. Then,
H-value is defined as,      
      
\begin{equation}      
\label{f2}      
H_{I,J}(i,j) = \frac{1}{|I||J|} {\sum_{i=0,j=0}}(RS_{I,J}(i,j)^2)      
\end{equation}   

{\it H-value} is a good measure and in many research papers and is used for
comparisons and it is also assigned as optimization goal~\cite{CC00,citeulike:2259666,
citeulike:6076707,bivusu2,CM09}. Furthermore, in \cite{citeulike:6370191},
they changed this optimization goal by adding multiplication
of $W(i,j) \times \theta$ where $W(i,j)$ is $1$ if $i$th row
and $j$th column is selected, otherwise, it is $0$, and $\theta$
is the overlap cost. With this extension, they claim
that overlapping biclusters would occur with
less probability due to the defined multiplication
as an overlap penalty.

In addition, rather than H-value, it is possible to use another metric called as {\it Hv-value}~\cite{KP07}      
{\it Hv-value} is similar to {\it H-value} and they claim that biclusters with similar      
row averages should get similar score values rather than {\it H-value}s of the same biclusters.
So they defined the {\it Hv-value} equation as,   
  
\begin{equation}      
\label{f6}      
Hv_{I,J}(i,j) = \frac{{\sum_{i=0,j=0}}((A_{i,j} - A_{Ij} - A_{iJ} + A_{IJ})^2)}{{\sum_{i=0,j=0}(A_{i,j} - A_{iJ})^2}}      
\end{equation}      

{\it Hv-value} is only used in its original paper, but it may be an alternative of {\it H-value} comparisons.

{\it Average Correlation Value}(ACV)~\cite{LW06}, is another variant of
{\it H-value}. According to the authors, it gives more desirable values for both
additive and multiplicative biclusters and in their setup for comparison with
{\it H-value} shows that ACV is more appropriate.
The equation of ACV score is,

\begin{eqnarray}
\begin{array}{rr}
ACV = max \Bigg\{ &
\begin{array}{r}
 \Big(\sum_{i}^{n}\sum_{j}^{n}(|R(i,j)|-n)\Big)/(n^2-n), \\
 \Big(\sum_{k}^{m}\sum_{l}^{m}(|C(k,l)|-m)\Big)/(m^2-m)
\end{array}
\end{array}
\end{eqnarray}

where $R$ is the function of the correlation coefficient between given two pair
indexes of rows and $C$ is the function of the correlation coefficient between
given two pair indexes of columns.

{\it P-value} is another metric for comparisons of biclusters. It is a scoring
function for enrichment measure that also implies the quality of
bicluster in terms of biological validation. Gene Ontology Consortium~\cite{citeulike:5044135}
defined the naming convention for each organism and genes of each organism
are appended to these categories. {\it P-value} maintains us to determine
the significance of genes of biclusters with respect to each participated GO category.
FuncAssociate~\cite{funcassociate}, GOTermFinder~\cite{citeulike:136570}
and several other tools are available for automated calculation of p-values and these
tools are not only available for biclustering analysis but also available for
other analyses~\cite{citeulike:4096566}. Rather than using GO categories and complicated tools,
in \cite{KP06,CM09} they calculate their own enrichment while using available
datasets where the general categories of each gene of the specific organisms
are determined. Using these datasets, they can calculate their own enrichment ratios
by looking at the highest representative biclusters of each general category.
This also refers to the enrichment and is an alternative to {\it P-value}.

Additionally, {\it Pearson Correlation Coefficient}(PCC) Scoring,
that is pairwise function and it is
similar to the calculation of {\it H-value} and varies between $[-1,1]$. 
PCC provides a linear correlation between a selected pair of genes
when the scores are approaching $1$ or $-1$. In the case of $0$
PPC score, the correlation is not linear.
In \cite{citeulike:5739089}, they use PPC to
collect correlated genes while maintaining good correlation. Also,
in \cite{citeulike:1356776}, they test with PCC to evaluate the performance of
their biclusters.

Finally, it is also meaningful to compare row means, columns means,
variance, and some other basic scoring schemes for each biclustering
result. Several biclustering papers
provide these comparisons as a supplement to the above scoring metrics.

In this section, we provide the definition of several scoring functions
that are used for the analysis of biclusters.
In general, {\it H-value} gives larger intervals. Hv-value gives narrow boundaries
and in some cases, the difference may not be significant. ACV and PCC are good
alternatives to {\it H-value}. {\it P-value} is also one of the main comparison metrics.
Among these, rather than {\it P-value}, none of them gives biological validation,
hence these scoring functions will not indicate the best biclusters in terms
of their biological value although they have the quality in scoring, this
does not mean that resulting biclusters are invaluable.

\subsubsection{Analysis with Biological Validation}

We review the computational analysis methodologies for biclusters. In this subsection,
we partially give biological validation methodologies. Integrating the
biclustering algorithms with biological validation or testing the resulting
biclusters by referring to biological data is important since we can show the biological importance
of biclusters rather than scoring as a metric. Last subsection, we mention about {\it P-value}. It provides
scores but these scores have also biological meaning. {\it P-value} relies
on GO categories with a set of genes. GO does the naming for categories
and has a predetermined set of genes for each category. For a bicluster consisting of
a set of genes and conditions, we consider genes such that in which
categories these genes are represented more, in other words, what portion
of bicluster genes belong to the selected categories. Two
widely used tools are freely available; FuncAssociate and GOTermFinder.
You can also write your own evaluation, by downloading
GO files from http://www.geneontology.org/. Therefore, P-value is
used in several research papers for comparisons due
to availability~\cite{citeulike:6076707,KP07,
citeulike:1356776,PBZW06,citeulike:2762934,bivusu2,
19208127,nsNMF,citeulike:2714170,citeulike:5649168}.

We have also another biological metric that is used for the validation
of biclusters. In \cite{PBZW06,XL2007,CM09}, they provide similar experiments
to validate their biclustering methods. This validation is based on
Protein-Protein Interaction(PPI) Networks. Their claim is the
correlation between genes tends to have more interacting PPI subnetworks.
This claim leads to some comparison metrics. You can use the reliability values of interactions
to measure the total reliability. You can also measure the average distance
between genes of biclusters in the original PPI network~\cite{PBZW06,XL2007}. In addition,
you can form a complete graph between genes of biclusters and check that for each
gene pair is there any interaction in the original PPI network~\cite{CM09}.

Finally, if we obtain some pre-knowledge about our input,
we can integrate this knowledge to validate biclusters. For instance,
{\it Colon Cancer Dataset} from \cite{citeulike:165055} is used
in \cite{MK03,XL2007}. It has 40 colon samples with tumor and
22 healthy colon samples, and approximately 6500 genes. In \cite{MK03,XL2007},
they tested their algorithm such that how many samples from each bicluster
have the cancer samples, with respect to set of genes. This experiment detects a
clue for colon cancer if a bicluster has a high ratio of tumor samples
and the resulting genes of this bicluster may be effective on colon cancer.
That is surely one of the important analyses of biclustering results and
with more available instances for other datasets, it would be good validation
of the biclusters.

We finish our review on visualization techniques and analysis methods.
Now, we give the related tools for biclustering that include at least
one of the mentioned visualization techniques.

\section{Survey of Existing Tools} 

In this section, we aim to review features of the existing tools
and the visualization and the analysis methods are outlined here.

\begin{table*}
\footnotesize
\caption{Overview of existing tools}
\label{overview} 
\begin{center}
\begin{tabular}[t]{lccp{3.0cm}p{3.0cm}p{2cm}p{1.2cm}}
\cline{1-7}
\multicolumn{1}{l}{}  & Heatmap & PC & Specialized Approach & Algorithm Support & Biological Evaluation & Bicluster Analysis \\ \cline{1-7}
\multicolumn{1}{l}{Expander 2003} & Yes & No & No & SAMBA & Yes  & Yes \\ 
\multicolumn{1}{l}{BicAT 2006} & Yes  & Yes & No & OPSM, Bimax, CC, XMOTIF, ISA & No  & Yes \\ 
\multicolumn{1}{l}{BiVoc 2006} & Yes & No & New Heatmap to visualize all biclusters & Import interface for existing results & No & No \\ 
\multicolumn{1}{l}{BiVisu 2007} & Yes & Yes & Modified PC plots & Own Algorithm & No  & Yes \\ 
\multicolumn{1}{l}{Bicoverlapper 2008} & Yes  & Yes & Force Directed Layouts & Import interface for existing results & Yes & Yes \\ 
\multicolumn{1}{l}{BiGGEsTS 2009} & Yes & Yes & Functional Category Tree View & ECCC & Yes & Yes \\ 
\multicolumn{1}{l}{IntegratedViz 2010} & No & No & Integrated Visualization with PPI networks & Import interface for existing results, REAL, Bimax, CC & Yes & Yes \\ 
\multicolumn{1}{l}{Robinviz 2011} & Yes  & Ye & A Graph Based, Reliability Measurement Criteria added upon IntegratedVis and Gene Ontology categories integrated to the results of biclustering algorithms & REAL, Bimax, CC & Yes & Yes \\ \hline

\end{tabular}
\end{center}
\end{table*}

In Table~\ref{overview}, we give an overview of visualization tools.
Table \ref{overview} supports our claim that there is fresh interest
in the topic. Several tools with different properties are in the literature.
Rather than visualization tools, there are also tools proposed for the self-execution of some algorithms~\cite{nsNMF,XL2007,SWPN09,epub3293}.Indeed, we go over specified tools in Table \ref{overview}
and we give the pros and cons of each tool and available
options.

\subsection{Expander}

Expander~\cite{Sharan03clickand,TSS02} is the eldest tool in our review.
Expander is a complete tool with its support for heatmap and PC visualizations, analysis, and
execution of algorithms. The tool supports the execution of their own biclustering
algorithm SAMBA and clustering algorithm CLICK. Expander also provides the visualization
of heatmaps, the resulting bitmaps can be saved easily, and biological evaluation
in terms of functional categories in GO using corrected P-value is available.
Obtained histogram of biclusters gives the specific encountered categories.
Hence this property supports the biclustering results.
It has other options such as Principal Component Analysis
and viewing Box Plots. Additionally, Expander allows saving sessions
which is also useful to continue at the point saved. It is coded in JAVA
and available at
\footnote[1]{Expander: \url{http://acgt.cs.tau.ac.il/expander/}}. 

\subsection{Biclustering Analysis Toolbox(BicAT)}

BicAT~\cite{bicat} is one of the early integrated tools for both execution, analyzing, and
visualization of biclusters. It supports heatmap and PC visualizations. The main
contribution of the tool is to provide a framework for
the execution of well-cited algorithms CC~\cite{CC00}, OPSM~\cite{BCKY02},
ISA~\cite{BIB03} and XMOTIF~\cite{MK03} and Bimax~\cite{PBZW06}. On the other
hand, the tool has no biological supported visualizations and narrow analysis
support. In general, it is one of the
earliest tools for bicluster visualization and it is famous because of its
variety of supported algorithms and simple GUI. BicAT is coded in JAVA and it
is free and downloadable at \footnote[2]{BicAT: \url{https://sop.tik.ee.ethz.ch/bicat/}}. 
However, the source code of the supported algorithms is closed.

\subsection{BiVoc}

BiVoc~\cite{bivoc} specializes in the heatmap representation by extending the representation
as multiple biclusters over the input matrix. The algorithm based on {\it PQ-Trees} is explained
in the previous section. It is innovative since there is no such work to show overlapping biclusters
as one heatmap for biclusters. The support of tools is limited due to specialization. They are supporting
a defined input format for submitting biclusters and navigation via a web-based interface.
Their main concern is to visualize the overlapping biclusters.
The methodology of BiVoc does not concentrate on having a unique label. One label
in the resulting heatmap could be represented several times but they claim that this repetition is
minimized with the guaranteed algorithm. Therefore, for less number of biclusters,
their method gives fine results in terms of the total number of rows
and columns on the layout. Vice versa, despite the minimization,
when the number of biclusters is high, their method gives several
rows and columns that may disturb the overall view.
Their solution to this problem is providing a web-based interface to follow and track
the results. The program is coded in C++ and freely available at 
\footnote[3]{BiVoc: \url{https://bioinformatics.cs.vt.edu/~murali/software/biorithm/bivoc.html}}. 
 
\subsection{BiVisu}

BiVisu~\cite{bivusu} is proposed with the algorithm called PM. Their contribution
is mainly on the PC plot drawing. Their approach
draws parallel coordinate plots by giving
plots of bicluster genes within plots of all genes inside the data. The color plots
correspond to genes of bicluster with a different color.
This enables the user to see the global view of the
parallel gene plots for the corresponding bicluster with
respect to a set of conditions in the bicluster. The tool
also provides a heatmap view. Further analysis of biclusters
are available in its GUI such as {\it H-value} and {\it Average Correlation Value}.
Although they propose an extension to PC plots, the view
has some problems of scaling as shown in ~\cite{bicoverlapper2}.
Their program is implemented in MATLAB and available at
\footnote[4]{BiVisu: \url{http://www.eie.polyu.edu.hk/~nflaw/Biclustering/}}. 

\subsection{Bicoverlapper}

BicOverlapper~\cite{bicoverlapper,bicoverlapper2} is one of the sophisticated tools
that mainly concentrates on the visualization techniques that exist before.
They also propose a brand new method. In this new method, they use
the force-directed layout of a graph of corresponding biclusters. The detail of the method is
given in the previous section. Since it provides visualization for
several biclusters, their method differs from other visualization methods
except BiVoc and IntegratedViz do. Their main contribution is handling several overlapping and nonoverlapping biclusters with given biological relevance
with respect to Transcriptional Regulatory Networks(TRN). They also support their main layout with
heatmaps and PC plots as evidence of their integrated
visualization, and they propose the 2D Bubble Map method by applying
from 3D version named Mountain Map visualization. Moreover, they
do not have implemented algorithms inside their tool,
however, they provide the import interface for the results of
biclustering algorithms. Their main concern is
the execution time of force-directed layouts. These layouts are simple
and easy to apply when the graph has a countable number of nodes.
In the case of biclustering results with higher dimensions,
meaning that many nodes, force-directed graph may slow the
execution and meaning of the main layout may be intervened.
Their tool is available at \footnote[5]{Bicoverlapper: \url{http://carpex.usal.es/~visusal/bicoverlapper/}} 
and they provided some example biological analysis in \cite{bicoverlapper2}.

\subsection{BiGGEsTS}

BiGGEsTS~\cite{biggest} is a recently proposed tool that supports both several visualizations
and analysis methods for biclusters. The tool also maintains the execution of their algorithm~\cite{MA09}.
Its GUI is similar to BicAT and thus user-friendly. They provide embedded visualization methods which
are heatmap, PC, multiple PCs, and enrichment tree visualization based on the method
described in the last section. BiGGEsTS maintains an integrated environment. You can obtain
both heatmaps and PCs. Also, their innovative enrichment tree visualization methodology gives
the biological significance of biclusters. Furthermore, in their multiple PC plots, they
also give the expression patterns that simplify the congested plots.
These simplified plots show the general trend of PC plots and the pattern of bicluster as well. 
Additionally, BiGGEsTS use the Graphviz tool to support their enrichment tree visualization.
The whole procedure except the production of graphs is executed
using Graphviz's dot program. BiGGEsTS supports sessions to save all the work
that is done during the session and also allows execution of their own algorithm eCCC.
Finally, BiGGEsTS is implemented using JAVA and it is under GPL license
available at \footnote[6]{BiGGEsTS: \url{http://kdbio.inesc-id.pt/software/biggests/} }.

\subsection{IntegratedViz and Robinviz}

IntegratedViz is an integrated visualization
tool and proposes an innovative approach of evaluation and validation of biclustering results
using biological data. IntegratedViz also supports Heatmap and PC plots visualizations. Particularly,
IntegratedViz concentrates on both global and local visualization
of biclusters. Global view shows each bicluster as a node
of weighted hierarchical layout and peripheral graphs corresponding to these graphs are accessible
via clicking. The details of the methodology are described in the previous section. Due to the nature of the
proposed methodology, IntegratedViz also provides some pieces of evidence for analysis such as scoring and enrichment
value of biclusters. Furthermore, their visualization method is
also supported with visual clues such as coloring of categories at
peripheral graphs which shows the main category of genes among pre-determined
ones, edge thickness, and node sizes. The main graph shows the H-values of each bicluster by arranging
the size of nodes. They also allow importing the results of algorithms and execution of given
algorithms in Table~\ref{overview}. The one disadvantage of the tool is again the problem of several biclusters
as it happens in BicOverlapper and BiVoc. To prevent visualization disturbance,
they added some scaling and hiding methods. IntegratedViz is written
in C++ and freely available at \footnote[7]{IntegratedViz: \url{http://webprs.khas.edu.tr/~cesim/}}
They also extended this tool and renamed as Robinviz
(Reliability Oriented Bioinformatics Network Visualization)~\cite{emre2011}.\footnote[8]{Robinviz: \url{https://github.com/aladagemre/robinviz}}. 

\section{Future Directions and Conclusion}

Indeed, we give the review of existing visualization and analysis methods providing a short explanation to each approach. We explain the common and recently proposed tools, mainly specialized in visualization, and have support for the described analysis options.

The problem of the biclustering area is the lack of fair analyzing tools and
specialized visualizations. Inevitably, there are many algorithms based on a variety of
disciplines due to the attraction of the topic but the dilemma is how to
analyze and show the abstract results to the end-users such as biologists.
Validating the visualizations with other biological data such as Functional Categories, PPIs and TRNs
surely be important in next-generation bicluster visualizations and analyses.
Heatmaps and Parallel Coordinate Plots are helpful but they are not able to
demonstrate the quality due to biological relevance. Since gene expression data
are the main input of biclustering algorithms, the validation of the
existing algorithms is surely done with the biological relevance of genes.
In that case, visualization methods should be able to show some clues
about these relevance to inform end-users. In \cite{citeulike:5917016},
they have a claim that networks in biology can appear complex and difficult
to decipher. They provide a concept in order to analyze these networks
using frequently employed visualization and analysis patterns.
This approach supports our claim for the bicluster visualization. The increasing amount of
data in subtopics of bioinformatics may result in relation
to the biclustering problem. As a result, the visualization approaches
from these topics can be integrated to decipher these relations.

We will also need to mention some standards. By now, we have some
standards for gene expression data and each tool or algorithm can easily
adopt these data, but we do not have the standard for importing the results
of biclusters. This is important since the proposed tools are based
on the assumption of having determined input. Tools have no chance to
follow many different formats so indeed the proposers of algorithms
should provide a stable format. This may be a new mark-up language format
that is easier to parse. Also, it is possible to use simple notation
as shown inside Figure~\ref{show}.

This notation provides the name of the algorithm, number of biclusters,
size of genes and conditions at each bicluster, and the labels of genes
and conditions in a bicluster. After these vital entities,
in the end, algorithm proposers may extend their file by adding their
scores if it is used and other information related to the
proposed algorithm.

Furthermore, providing an open-source codes or self-executable
programs for the proposed biclustering algorithms should be beneficial.
Therefore, it is possible to look at the other proposed method without
reimplementation. Especially in biclustering, it is more desirable
to have open-source codes or self-executable
programs to help the prospective authors for their comparisons.

\begin{figure*}[t]	   
\begin{center}	   

\begin{lstlisting}
\\ Sample Result File
\\ Algorithm: CC
BiclusterNumber: 2
7 3
gene1 gene5 gene11 gene12 gene14 gene15 gene16
cond1 cond3 cond5
3 5
gene2 gene3 gene4
cond2 cond4 cond6 cond7 cond8
\end{lstlisting} 
\caption{\sf Suggested Bicluster Result Output}	   
\label{show}	   
\end{center}	   
\end{figure*}

In a conclusion, we believe visualization and analysis of biclusters are
still hot topics. With this review paper, we imply the analysis
and visualization methods and we give the future directions.
Understanding and analyzing biclustering results,
setting up bridges to other related topics and providing a
visualization approaches by integration with other topics would
be soon more important. Hence extending the topic to the other
applications would be easier.

\section*{Acknowledgements}

Melih S\"{o}zdinler is paid by The Scientific and Technological Research Council of Turkey(TUBITAK)[BIDEB-2211]. Additional thanks to my PhD advisors; Can \"{O}zturan, Turkan Haliloglu, Arzucan \"{O}zg\"{u}r and Cesim Erten

{\it Conflict of Interest:} None declared.

\bibliographystyle{unsrtnat}
\bibliography{references}

\end{document}